\documentclass[preprintnumbers,prl,twocolumn,floatfix,preprintnumbers,amsmath,amssymb,nofootinbib,superscriptaddress]{revtex4}
\usepackage{graphicx}
\usepackage{enumitem}
\usepackage{epsfig}
\usepackage{bm}
\usepackage{amsfonts}
\usepackage{color}
\usepackage{subfigure}
\usepackage{amsmath}
\usepackage{amssymb}
\usepackage{graphicx, comment}
\usepackage{makecell}
\usepackage{float}
\usepackage[normalem]{ulem}
\usepackage{xcolor}

\usepackage[utf8]{inputenc}
\usepackage[english]{babel}

\usepackage[
      colorlinks=true,
      linkcolor=blue,
      urlcolor=magenta,
      filecolor=green,
      citecolor=red,
      pdfstartview=FitV,
      pdftitle={},
        pdfauthor={Koushik Dutta, Anirban Roy, Ruchika, Anjan Sen, M.M. Sheikh-Jabbari},
        pdfsubject={},
        pdfkeywords={},
        pdfpagemode=None,
        bookmarksopen=true
      ]{hyperref}
\usepackage{color}

\newcommand{\bcm}{}

\usepackage{ulem,xpatch}

\usepackage{amsmath,amssymb}

\begin{document}

\preprint{IPM/P-2019/nnnn}

\title{Cosmology With Low-Redshift Observations: No Signal For New Physics}

\author{Koushik Dutta}
\email{koushik.physics@gmail.com}
\affiliation{Theory Division, Saha Institute of Nuclear Physics, HBNI, 1/AF Salt Lake, Kolkata-700064, India}
\author{Anirban Roy}%
\email{aroy@sissa.it}
\affiliation{SISSA, Via Bonomea 265, 34136, Trieste, Italy}

\author{Ruchika}
\email{ruchika@ctp-jamia.res.in}
\author{Anjan A Sen}
\email{aasen@jmi.ac.in}
\affiliation{Centre for Theoretical Physics, Jamia Millia Islamia, New Delhi-110025, India.}%
\author{M.~M.~Sheikh-Jabbari}
\email{jabbari@theory.ipm.ac.ir}
\affiliation{School of physics, Inst. for research in fundamental sciences (IPM), P.O.Box 19395-5531, Tehran, Iran}
\affiliation{The Abdus Salam ICTP, Strada Costiera 11, 34151 Trieste, Italy}%

\date{\today}

\begin{abstract}
We analyse various low-redshift cosmological data from Type-Ia Supernova, Baryon Acoustic Oscillations, Time-Delay measurements using Strong-Lensing, $H(z)$ measurements using Cosmic Chronometers and growth measurements from large scale structure observations for $\Lambda$CDM and some different dark energy models. By calculating the Bayesian Evidence for different dark energy models, we find out that the $\Lambda$CDM  still gives the best fit to the data with  $H_{0}=70.3^{+1.36}_{-1.35}$ Km/s/Mpc (at $1\sigma$). This value is in  $2\sigma$ or less tension with various low and high redshift measurements for $H_{0}$ including SH0ES, Planck-2018 and the recent results from H0LiCOW-XIII. The derived constraint on $S_{8}=\sigma_{8}\sqrt{{\Omega_{m0}}/{0.3}}$ from our analysis is $S_{8} = 0.76^{+0.03}_{-0.03}$, fully consistent with direct measurement of $S_{8}$ by KiDS+VIKING-450+DES1 survey. We hence conclude that the $\Lambda$CDM model with parameter constraints obtained in this work is consistent with different early and late Universe observations within $2\sigma$. We therefore, do not find any compelling reason to go beyond concordance $\Lambda$CDM model. 

\end{abstract}

\maketitle

\section{Introduction}
Different cosmological observations of the last couple of decades have surprised us with extraordinary results and puzzles, in particular the fact that our universe is now in an accelerated expansion, first confirmed by Type-Ia Supernova observations (SN) \cite{Perlmutter:1998np,Riess:1998cb} and later verified by host of other cosmological observations including the CMB  \cite{Komatsu:2010fb,Ade:2013zuv} and LSS observations \cite{Tegmark:2003ud}. The evidence for the accelerated expansion requires existence of a repulsive gravity at large cosmological scales defying our usual notion of gravity as an attractive force. This demands either the existence of mysterious dark energy in the energy budget of the universe or the modification of Einstein gravity at large cosmological scales. A positive cosmological constant ($\Lambda$) is the simplest way to incorporate repulsive gravity in Einstein's GR. The concordance $\Lambda$CDM model is consistent with most of the cosmological observations today including the latest Planck-2018  \cite{Aghanim:2018eyx} measurements of the Cosmic Microwave Background (CMB). Nonetheless, cosmological constant as dark energy suffers from severe technical naturalness and from fine-tuning of the observed value of $\Lambda$, motivating people to consider dark energy models beyond the cosmological constant.

The first serious tension for $\Lambda$CDM model emerges from the Local measurement of the expansion rate of the universe (SH0ES project) by Riess et al. in 2016 (R16) \cite{Riess:2016jrr} and then the latest one in 2019 (R19)\cite{Riess:2019cxk}. Their model independent measurement of the Hubble parameter at present is $H_{0} = 74.03 \pm 1.42$ Km/s/Mpc which is in $4.4\sigma$ tension with the result $H_{0} = 67.4 \pm 0.5$ Km/s/Mpc from the latest Planck-2018  \cite{Aghanim:2018eyx} compilation for a concordance $\Lambda$CDM model. If one also combines the latest Strong-Lensing time-delay measurements of $H_{0}$ by H0LiCOW (H0LiCOW-XIII) \cite{Wong:2019kwg} to the local measurement of $H_{0}$ by R19, the tension with Planck-2018 value of $H_{0}$ exceeds $5\sigma$ level for the $\Lambda$CDM model \cite{Verde:2019ivm}.

The effects of accelerated expansion and dark energy are also imprinted in the Baryon Acoustic Oscillations (BAO) at the time of recombination. The observations of BAO peaks in galaxy surveys give accurate measurements of $r_{d}H_{0}$, the product of sound horizon size at drag epoch $r_{d}$ and the present day Hubble parameter $H_{0}$. For the concordance $\Lambda$CDM model, CMB measurement by Planck-2018 gives $r_{d} = 147.05 \pm 0.30$ Mpc  \cite{Aghanim:2018eyx}. Using this value of $r_{d}$, inverse distance ladder method \cite{Aubourg:2014yra} using BAO and CMB, gives $H_{0}$ around $67$ Km/s/Mpc, which is in more than $4\sigma$ tension with R19. On the other hand, assuming low-redshift anchor for $H_{0}$ from R16 and H0LiCow, BAO measurements gives $r_d$ around $136$ Mpc  \cite{Evslin:2017qdn} irrespective of the dark energy model. This is in $3\sigma$ tension with value obtained by Planck-2018 for $\Lambda$CDM. While the amount of tension depends on how one calibrates BAO measurements, there is always a large tension between high and low redshift observations either for $H_{0}$ or for $r_{d}$.

This large tension may be due to unaccounted systematic in the local $H_{0}$ measurements. Whether we are living in a local underdense region or a void as possible systematic has been studied  \cite{Wu:2017fpr,Kenworthy:2019qwq} (see however \cite{Colgain:2019pck}), showing that this effect is too small to resolve the discrepancy of $6$ Km/s/Mpc in $H_{0}$ between low and high redshift measurements  \cite{Boehringer:2019xmx}. Similarly, weak lensing of the Supernova magnitudes has been shown to have too little effect to account for the current tension \cite{Smith:2013bha}.

Efforts have also been made to introduce new physics both at early time as well as at late times to resolve this tension. Models with early dark energy \cite{Poulin:2018cxd}, extra relativistic species \cite{Wyman:2013lza}, decay of dark matter to dark radiation \cite{ Pandey:2019plg, Vattis:2019efj}, evolving dark energy \cite{Zhao:2017cud,Zhang:2017idq, DiValentino:2017zyq, Sola:2018sjf, DiValentino:2019exe}, interacting dark energy \cite{Agrawal:2019dlm}, possibility of existence of negative cosmological constant \cite{Dutta:2018vmq, Visinelli:2019qqu}, phantom dark energy \citep{Li:2019yem} , effect of non-Gaussian primordial fluctuation \cite{Adhikari:2019fvb}, existence of screened fifth force in the local universe \cite{Desmond:2019ygn} or modified gravity theory at low redshifts \cite{Kazantzidis:2019dvk} are some of the recent attempts to resolve this tension.

Recently an independent local measurement of the $H_{0}$  has been carried out by the Carnegie-Chicago Hubble Program (CCHP) and presented by Freedman et al (F19) \cite{Freedman:2019jwv}.  Instead of the calibration using the Cepheids to Type-Ia Supernova magnitudes, as used in R19 measurements, they have used the calibration of the Tip of the Red Giant Branch (TRGB) in the nearby galaxies. With this, F19 presented the measured local $H_{0}$ value which is $H_{0} = 69.8 \pm 0.8$ (stat)$\pm 1.7$(sys) Km/s/Mpc. This agrees with the Planck-2018 measured value within $1.2\sigma$ and with R19 measured value within $1.7\sigma$. Given that the method of calibration in R19 and F19 are completely independent, the comparatively lower value obtained by F19 shows that one needs to be careful before drawing any conclusion on  the tension between local measurement of $H_{0}$ and that extracted from the CMB. Recently, using both TRGB and SNe Ia distance ladder, Yaun et al. have determined the value for the Hubble constant to be $H_0 = 72.4 \pm 1.9$ Km/S/Mpc \cite{Yuan:2019npk}.

As mentioned above, the recent results by H0LiCOW-XIII \cite{Wong:2019kwg} experiment for time-delay measurements with six Strong Lensing systems, have reported an independent measurement of  $H_{0}$ for  $\Lambda$CDM model which is $H_{0} = 73.3^{+1.7}_{-1.8}$ Km/s/Mpc. This value is consistent with both R19 and F19 results but is in $3.1\sigma$ tension with Planck-2018 measured value. However, the measured values of $H_{0}$ for individual six lenses vary from $68.9$ Km/s/Mpc to $81.1$ Km/s/Mpc, too wide for any definite conclusion regarding the $H_{0}$ tension.

With this background, we reanalyze the available low-redshift observational data assuming different dark energy models which affects only the late time cosmology. We exclude the local measurements of $H_{0}$ either by R19 or F19 so that the measured values of different cosmological parameters including $H_{0}$, are not biased due to different calibrations used in R19 and F19. While using the BAO data, we also do not bias our results by assuming a prior on $r_{d}$ (sound horizon scale at drag epoch) as given by Planck-2018. We allow it to be a free parameter in our analysis and determine it from low-redshift observations to corroborate with the findings from the Planck-2018 results. Our aim is to examine if low-redshift data (without local measurement of $H_{0}$) also exhibit significant tensions in measured values of different cosmological parameters. In the next two sections we respectively discuss the dark energy models/parameterizations and cosmological data that we use for our analysis. We then present our results. Finally, we conclude with some discussions that includes the implications of our results for $S_8$ observations.

\section{Modeling Late Time Cosmology}

To begin with, we assume spatially flat FRW cosmology for the background universe and normalise the present value of scale factor $a_{0} =1$. Throughout the manuscript, subscript ``0" represents parameter values at present $(z=0)$. We consider the following dark energy models for our background cosmology:
\begin{itemize}[leftmargin=*]
    \item {\bf $\Lambda$CDM model} with the Hubble parameter $H(z)$, 
    \begin{equation}
       \frac{H^{2}(z)}{H_{0}^2} = \Omega_{m0}(1+z)^3 + (1-\Omega_{m0}).
    \end{equation}
 Here $\Omega_{m0}$ is matter density parameter today.    
    \item {\bf $w$CDM model} with dark energy model of a constant equation of state $w$ and the Hubble parameter $H(z)$,  
    \begin{equation}
         \frac{H^{2}(z)}{H_{0}^2} = \Omega_{m0}(1+z)^3 + (1-\Omega_{m0})(1+z)^{3(1+w)}~.
    \end{equation}
    \item {\bf CPL model}: Dark energy model with an equation of state $w(z) = w_{0} + w_{a}(1-a)=w_{0} + w_{a}\frac{z}{1+z}$ where $w_{0}$ and $w_{a}$ are two arbitrary constants  \cite{Chevallier:2000qy,Linder:2002et} and
    \begin{equation}
     \frac{H^{2}(z)}{H_{0}^2} = \Omega_{m0}(1+z)^3 + (1-\Omega_{m0})f(z),
     \end{equation}
    where $f(z)=\exp\left(3\int^z\frac{(1+w(x))}{1+x}dx\right)$.
    \item {\bf Pade  Model for dark energy}: Unlike the previous cases we do not assume any specific parametrisation for the dark energy equation of state. Rather, we assume a generic dark energy behaviour beyond a cosmological constant $\Lambda$. The simplest way to model deviation from a constant dark energy density is Taylor-expansion around  $\Lambda$ for the dark energy density. As we are interested in the low-redshift observations, we do the expansion around the present day ($a_0=1$):
    \begin{equation}
        \rho_{de}(a) = \rho_{0} + \rho_{1} (1-a) + \rho_{2} (1-a)^2 + ...
    \end{equation}
Here $\rho_{0}$ is the present value of the dark energy density and $\rho_{1}$ and $\rho_{2}$ are two expansion parameters. If one rewrites the above expression in terms of the redshift $z$, the dark energy density takes the form:
    \begin{equation}
         \rho_{de}(a) = \frac{\rho_{0} + (2\rho_{0}+\rho_{1})z + (\rho_{2}+\rho_1+\rho_0) z^2}{1+2z+z^2}\nonumber
    \end{equation}
which is Pade expansion of order (2,2) around $z=0$ with properly identified expansion coefficients \cite{Saini:1999ba}. In our subsequent analysis, we assume the dark energy density as a general Pade expansion of order (2,2) around $z=0$. The Hubble parameter $H(z)$ is then given by
    \begin{equation}
         \frac{H^{2}(z)}{H_{0}^2} = \Omega_{m0}(1+z)^3 + (1-\Omega_{m0}){\cal P}(z)~,
    \end{equation}
where 
\begin{equation}
{\cal P}(z)=\frac{1+P_{1}z+P_{2}z^2}{1+Q_{1}z+Q_{2}z^2}~,
\end{equation}
with $P_{1}$, $P_{2}$, $Q_{1}$, and $Q_{2}$ being four expansion parameters. This parametrisation of dark energy density allows a host of dark energy behaviour including phantom to non-phantom crossing as well as a negative effective dark energy density which may arise in scalar tensor theory models of dark energy \cite{Dutta:2018vmq}. Note that for very small and very large redshifts, the dark energy density approaches a constant value as in cosmological constant and the model approaches $\Lambda$CDM.
\end{itemize}

\begin{table}[t]
\centering
\begin{tabular}{ccc}
\Xhline{\arrayrulewidth}
\textbf{Parameter} & \textbf{Models} & \textbf{Prior (uniform)}\\ 
\Xhline{\arrayrulewidth}
$\Omega_{m0}$ & All & [0.1, 0.9]\\
$H_{0}$ & All & [50, 90] Km/s/Mpc\\
$\sigma_8$ & All & [0.6, 1.0]\\
$r_{d}$ & All & [130,170]\\
$w$ & $wCDM$,  & [-1.9, -0.4]\\
$w_0$ & CPL &[-1.9, -0.4]\\
$w_a$ & CPL & [-4.0, 4.0]\\
\Xhline{\arrayrulewidth} 
\end{tabular} 
\caption{Parameters used in different models and their prior. }\label{Table-1}
\end{table}


\section{Data Sets Used}

To constrain different dark energy models mentioned above, we consider the following observational data:
\begin{itemize}[leftmargin=*]
    \item Isotropic BAO measurements by 6dF survey ($z=0.106$) \cite{Beutler:2011hx}, SDSS-MGS survey ($z=0.15$) \cite{Ross:2014qpa} as well as by eBOSS quasar clustering ($z {=}1.52$) \cite{Ata:2017dya}. We also consider anisotropic BAO measurements by BOSS-DR12 at $z=0.38,0.51,0.61$ \cite{Alam:2016hwk}. Finally we consider BAO measurement by BOSS-DR12 using Lyman-$\alpha$ samples at $z=2.4$ \cite{Bourboux:2017cbm}.
    {Hereafter, we will call all these data together as ``BAO" data}.
    
    \item Angular diameter distances measured using water megamasers under the Megamaser Cosmology Project at redshifts $z=0.0116, 0.0340, 0.0277$ for Megamasers UGC 3789, NGC 6264 and NGC 5765b respectively \cite{Reid:2008nm,Reid:2012hm,Kuo:2012hg,Gao:2015tqd}, hereafter ``MASERS". We will also make some comments on the very
    recent NGC 4258 data and its implications for $H_0$ \cite{Reid:2019tiq} in the discussion part.
    \item Strong lensing time-delay measurements by H0LiCOW experiment (TDSL) by Bonvin et al. (2017) \cite{Bonvin:2016crt} for three lens systems, hereafter hereafter “SL” data. Note that we have not considered the recent H0LiCOW-XIIIsurements \citep{Wong:2019kwg} of six lenses. We discuss this in the later part of the paper.
    
    \item 
    Latest Pantheon data for SNIa in terms of $H(z)/H_0$ as compiled by Gomez-Valent \& Amendola (2018) \citep{Gomez-Valent:2018hwc}, here- after “SN” data.
    
    \item
    OHD data for Hubble parameter as a function of redshift using cosmic chronometers as compiled in Gomez-Valent \& Amendola (2018) \citep{Gomez-Valent:2018hwc}, hereafter “CC” data.
    
    \item
     Finally, the measurements of $f\sigma_{8}$ by various galaxy surveys as compiled by Nesseris et al. \citep{Nesseris:2017vor}, hereafter ``$f\sigma_8$'' data. This is ``Gold-17" sample for $f\sigma_{8}$ measurements. We have implemented the corrections for the dependence on fiducial cosmology used to convert redshifts into distances as described in \citep{Nesseris:2017vor} (see also \cite{Kazantzidis:2018rnb}). Here $f$ is the growth parameter for the linear matter density fluctuation $f=\frac{d\log\delta}{d\log a}$, with $\delta$ being the linear matter overdensity and $\sigma_{8}$ is the {\it rms} fluctuation of linear matter overdensity on $8h^{-1}$ Mpc scale.

\end{itemize}
Combination of the first four set of  data {(BAO + MASERS + SL + SN)} is termed as ``BASE" set in our subsequent analysis. The priors for different parameters considered in our analysis are given in Table \ref{Table-1}. We use the publicly available code ``Emcee" \citep{ForemanMackey:2012ig} to generate the MCMC  for different models for the above set of data.

\begin{table}[t]
\centering
\begin{tabular}{ |p{2.7cm}||p{1.9cm}|p{1.7cm}|p{1.6cm}|  }
 \hline
 \multicolumn{4}{|c|}{$\Lambda$CDM} \\
 \hline
 Parameters& BASE &+CC&+$f\sigma_{8}$\\
 \hline
 $\Omega_{m0}$&$0.29^{+0.014}_{-0.014}$&$0.29^{+0.013}_{-0.013}$&$0.29^{+0.013}_{-0.013}$\\
 $r_{d}$ (Mpc)&$141.72^{+4.55}_{-4.56}$&$144.66^{+2.81}_{-2.83}$&$144.71^{+2.9}_{-2.9}$\\
 $H_{0}$ (Km/s/Mpc)&$71.74^{+2.22}_{-2.22}$ & $70.26^{+1.38}_{-1.37}$&$70.3^{+1.36}_{-1.35}$\\
 $\sigma_{8}$&\hspace*{7mm}---&\hspace*{7mm}---&$0.77^{+0.03}_{-0.03}$\\
 $S_{8}=\sigma_{8}\sqrt{{\Omega_{m0}}/{0.3}}$&\hspace*{7mm}---&\hspace*{7mm}---&$0.76^{+0.03}_{-0.03}$\\
 \hline
\end{tabular}
\caption{Constraints on parameters for $\Lambda$CDM for different set of observation data. The error bars quoted are at $1\sigma$ confidence interval. ``BASE" denotes ``BAO+MASERS+SL+SN".}\label{Table-2}
\end{table}

\begin{table}[t]
\centering
\begin{tabular}{ |p{2.7cm}||p{1.9cm}|p{1.7cm}|p{1.6cm}|  }
 \hline
 \multicolumn{4}{|c|}{$w$CDM} \\
 \hline
 Parameters& BASE &+CC&+$f\sigma_{8}$\\
 \hline
 $\Omega_{m0}$&$0.284^{+0.018}_{-0.018}$&$0.285^{+0.017}_{-0.017}$&$0.30^{+0.009}_{-0.009}$\\
 $r_{d}$ (Mpc)&$141.76^{+4.44}_{-4.48}$&$144.77^{+2.88}_{-2.87}$&$144.51^{+2.90}_{-2.92}$\\
 $H_{0}$ (Km/s/Mpc)&$71.44^{+2.21}_{-2.17}$ & $69.97^{+1.4}_{-1.4}$&$70.27^{+1.44}_{-1.43}$\\
 $w$&$-0.96^{+0.07}_{-0.06}$&$-0.96^{+0.06}_{-0.06}$&$-1.02^{+0.03}_{-0.03}$\\
 $\sigma_{8}$&\hspace*{7mm}---&\hspace*{7mm}---&$0.76^{+0.03}_{-0.03}$\\
 $S_{8}=\sigma_{8}\sqrt{{\Omega_{m0}}/{0.3}}$&\hspace*{7mm}---&\hspace*{7mm}---&$0.756^{+0.03}_{-0.03}$\\
 \hline
\end{tabular}
\caption{The same as Table \ref{Table-2} but for $w$CDM model.}\label{Table-3}
\end{table}

\begin{table}[t]
\centering\begin{tabular}{ |p{2.7cm}||p{1.9cm}|p{1.7cm}|p{1.6cm}|  }
 \hline
 \multicolumn{4}{|c|}{CPL} \\
 \hline
 Parameters& BASE &+CC&+$f\sigma_{8}$\\
 \hline
 $\Omega_{m0}$&$0.232^{+0.054}_{-0.063}$&$0.23^{+0.054}_{-0.06}$&$0.241^{+0.039}_{-0.041}$\\
 $r_{d}$ (Mpc)&$141.28^{+4.56}_{-4.63}$&$144.75^{+2.8}_{-2.73}$&$144.88^{+2.84}_{-2.95}$\\
 $H_{0}$ (Km/s/Mpc)&$71.87^{+2.33}_{-2.28}$ & $70.22^{+1.4}_{-1.4}$&$70.19^{+1.48}_{-1.46}$\\
 $w_{0}$&$-0.96^{+0.08}_{-0.08}$&$-0.96^{+0.08}_{-0.08}$&$-0.97^{+0.07}_{-0.07}$\\
 $w_{a}$&$0.56^{+0.42}_{-0.46}$&$0.57^{+0.41}_{-0.45}$&$0.56^{+0.38}_{-0.39}$\\
 $\sigma_{8}$&\hspace*{7mm}---&\hspace*{7mm}---&$0.85^{+0.08}_{-0.07}$\\
 $S_{8}=\sigma_{8}\sqrt{{\Omega_{m0}}/{0.3}}$&\hspace*{7mm}---&\hspace*{7mm}---&$0.76^{+0.029}_{-0.029}$\\
 \hline
\end{tabular}
\caption{The same as Table \ref{Table-2} but for CPL model.}\label{Table-4}
\end{table}

\begin{table}[t]
\centering
\begin{tabular}{ |p{2.7cm}||p{1.9cm}|p{1.7cm}|p{1.6cm}|}
 \hline
 \multicolumn{4}{|c|}{PADE} \\
 \hline
 Parameters& BASE &+CC&+$f\sigma_{8}$\\
 \hline
 $\Omega_{m0}$&$0.355^{+0.12}_{-0.12}$&$0.31^{+0.09}_{-0.07}$&$0.263^{+0.046}_{-0.044}$\\
 $r_{d}$ (Mpc)&$140.35^{+4.7}_{-4.81}$&$144.76^{+2.9}_{-2.9}$&$145.06^{+2.95}_{-2.94}$\\
 $H_{0}$ (Km/s/Mpc)&$73.6^{+2.7}_{-2.7}$ & $71.22^{+1.8}_{-1.8}$&$71.14^{+1.7}_{-1.7}$\\
 $P_{1}$&$-0.735^{+2.31}_{-1.92}$&$1.48^{+2.28}_{-2.13}$&$2.55^{+1.58}_{-1.59}$\\
 $P_{2}$&$-0.43^{+1.92}_{-1.83}$&$0.17^{+1.99}_{-1.65}$&$0.67^{+2.11}_{-2.05}$\\
 $Q_{1}$&$0.9^{+1.98}_{-1.2}$&$1.49^{+2.16}_{-1.78}$&$2.41^{+1.69}_{-1.77}$\\
 $Q_{2}$&$0.14^{+0.11}_{-0.11}$&$0.101^{+0.096}_{-0.098}$&$0.095^{0.09}_{-0.09}$\\
 $\sigma_{8}$&\hspace*{7mm}---&\hspace*{7mm}---&$0.700^{+0.03}_{-0.03}$\\
 $S_{8}=\sigma_{8}\sqrt{{\Omega_{m0}} / {0.3}}$&\hspace*{7mm}---&\hspace*{7mm}---&$0.654^{+0.06}_{-0.06}$\\
 \hline
\end{tabular}
\caption{The same as Table \ref{Table-2} but for Pade model.}\label{Table-5}
\end{table}

\section{Results}

\begin{figure}
\resizebox{250pt}{205pt}{\includegraphics{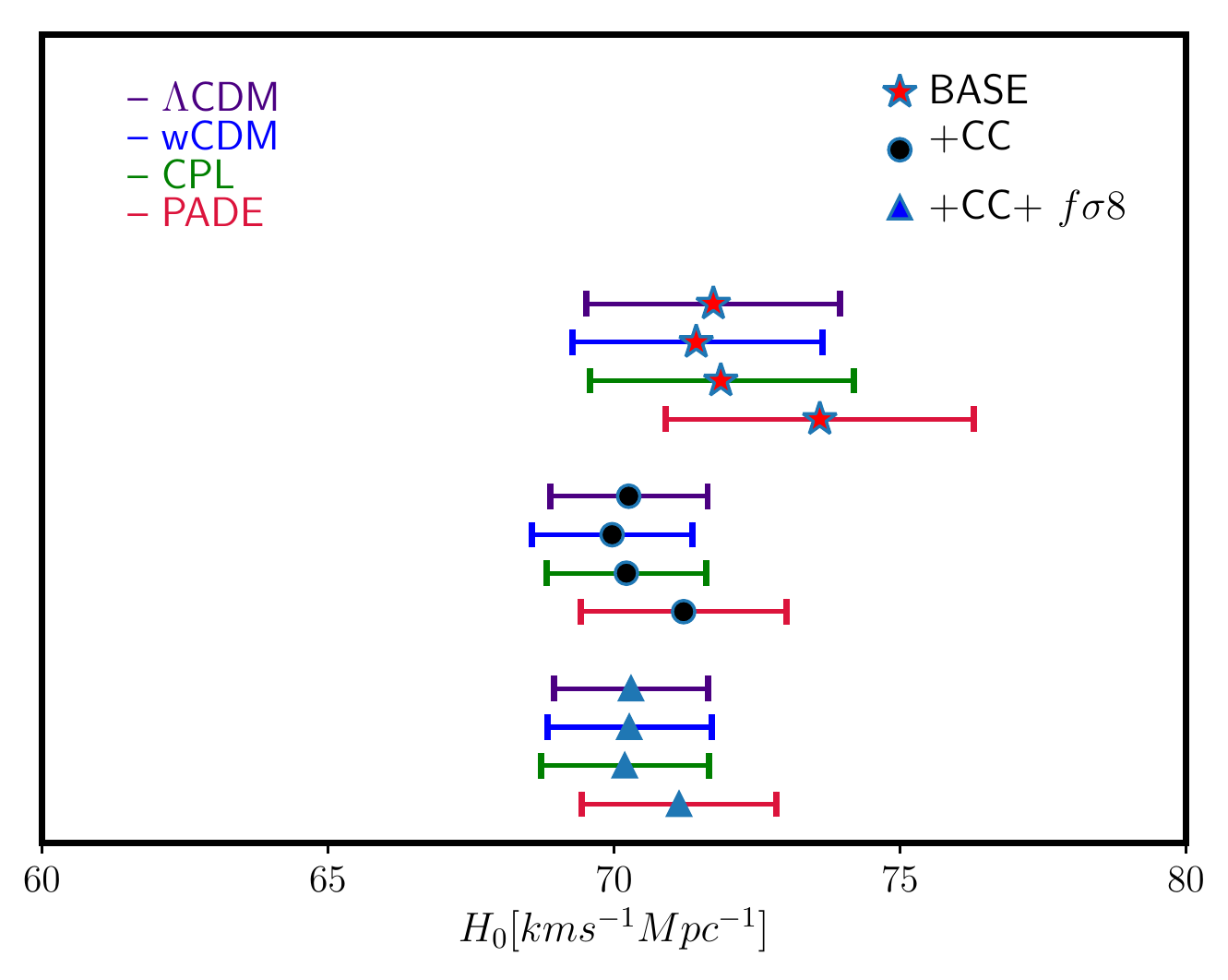}}
\caption{Values of $H_{0}$ with $1$-$\sigma$ error for several models studied in the work. The results are shown for different combinations of data-sets studied.}\label{figy:hdr} 
\end{figure}

The constraints on different cosmological as well as  model parameters for different combination of data, are shown in Tables \ref{Table-2}, \ref{Table-3}, \ref{Table-4} and \ref{Table-5}. Different columns in these tables correspond to different combinations of data sets where we separate out the effects of adding Cosmic Chronometers (CC) data, and the $f\sigma_8$ data. As one can see from these tables, the constraints on three cosmological parameters ($\Omega_{m0}, r_{d}, H_{0}$), are nearly the same for different dark energy models except that CPL and Pade prefer slightly lower value of $\Omega_{m0}$ and Pade prefers slightly higher value for $H_{0}$. But all of them are consistent within $1\sigma$ error bar. Note that the inclusion of Cosmic Chronometers data always make the $H_0$ value smaller which is more compatible with the derived $H_0$ value from the high red-shift data. Further inclusion of $f\sigma_{8}$ does not change $H_0$ value much. All the values of $H_{0}$ for different dark energy models for the full data sets (BASE+CC+$f\sigma_{8}$) are fully consistent with F19 \cite{Freedman:2019jwv}. For $\Lambda$CDM, we obtain $H_{0} = 70.3^{+1.36}_{-1.35}$ for full data sets. This is in tension with both Planck-2018 \cite{Aghanim:2018eyx} and R19 \cite{Riess:2019cxk} results at $2.01\sigma$ and $1.9\sigma$ respectively and with recent H0LiCOW-XIII measurements for six lenses at $1.33\sigma$. In other words, for $\Lambda$CDM, our $H_{0}$ measurement with various low-redshift observations are consistent with R19, F19, Planck-2018 as well as recent H0LiCOW-XIII \cite{Wong:2019kwg} measurements {with $2\sigma$ or less} error bar. Moreover this result still holds for our different dark energy models. Note that we use the H0LiCOW(2017) data for three lenses, however, our measured value for $H_{0}$ is consistent with the combined measurement for $H_{0}$ by latest H0LiCOW-XIII for six lenses. {Figure \ref{figy:hdr} depicts the marginalised $H_0$ value with $1$-$\sigma$ error for different models for different combinations of the data, clearly showing the consistency of $H_{0}$ measurements for different dark energy models.}

At this point, it is important to remind ourselves that the derived parameter values must not be in substantial conflict with the high red-shift Planck data. In our analysis, we do not take the full Planck likelihood in consideration to derive the best fit parameters. Rather, we cross-check that our derived parameters are not in major deviations from the Planck-2018 parameter values. As noted in the Introduction, CMB  measurements  by Planck-2018 gives the sound horizon scale at drag epoch $r_d= 147.05 \pm 0.30$ Mpc.
Using the low red-shifts data, we obtain $r_{d} = 144.68^{+2.9}_{-2.9}$ for $\Lambda$CDM model with all the data sets, which is in less than $1\sigma$ tension with Planck-2018 result. Here also, note the importance of Cosmic Chronometers data whose inclusion always make the $r_d$ value larger, making it more consistent with CMB result for $r_d$. 

\begin{figure}[t]
\resizebox{240pt}{180pt}{\includegraphics{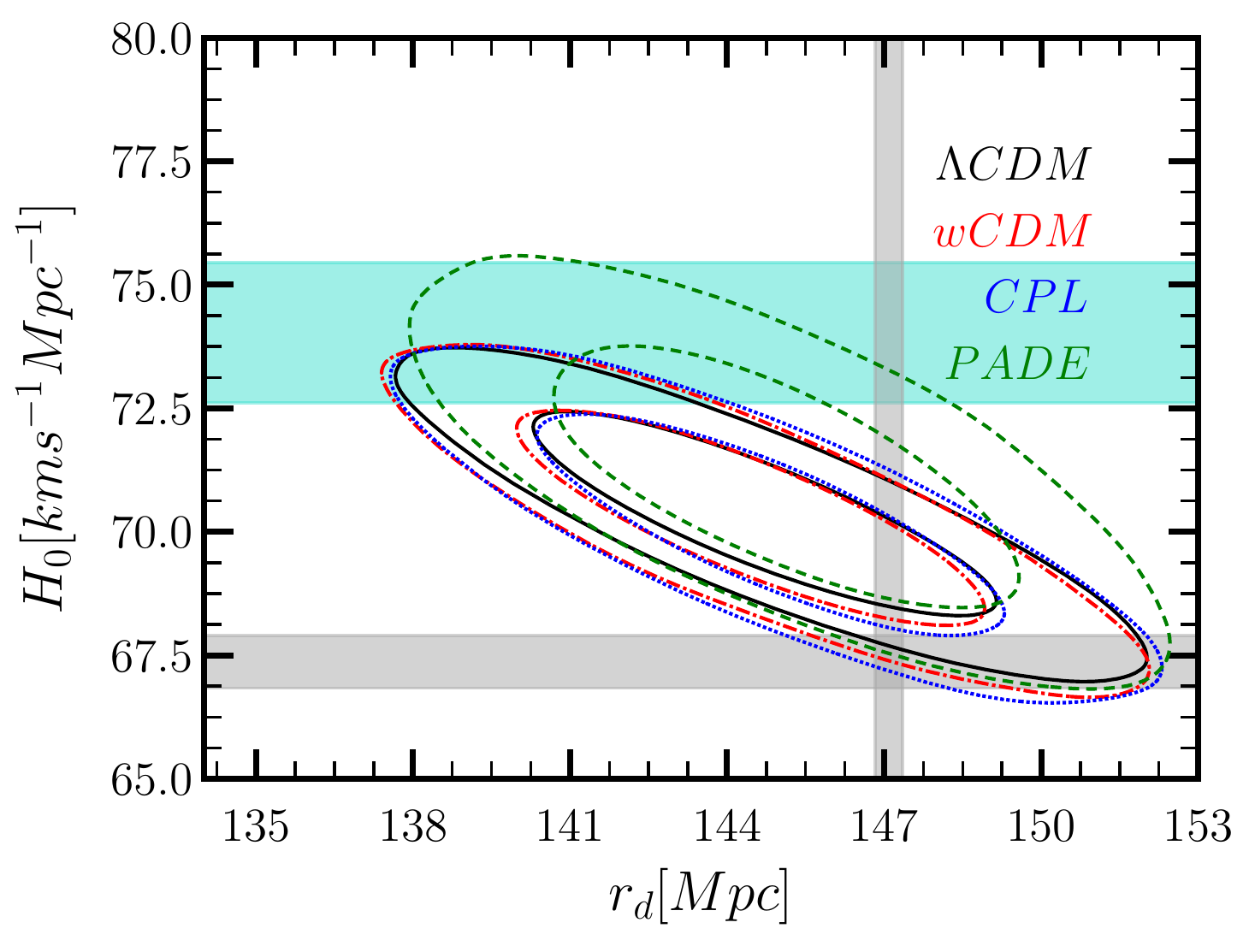}}
\caption{\label{fig:hdr}$1\sigma$ and $2\sigma$ constrained contours in $H_{0}-r_{d}$ parameter plane. The horizontal green band is $1\sigma$ constraints on $H_{0}$ by R19. The horizontal (vertical) grey band is constraint on $H_{0}$ (on $r_{d}$) from Planck-2018. Here the ``BASE+CC+$f\sigma_{8}$" data sets is used.  }
\end{figure}

\begin{figure}
\resizebox{250pt}{188pt}{\includegraphics{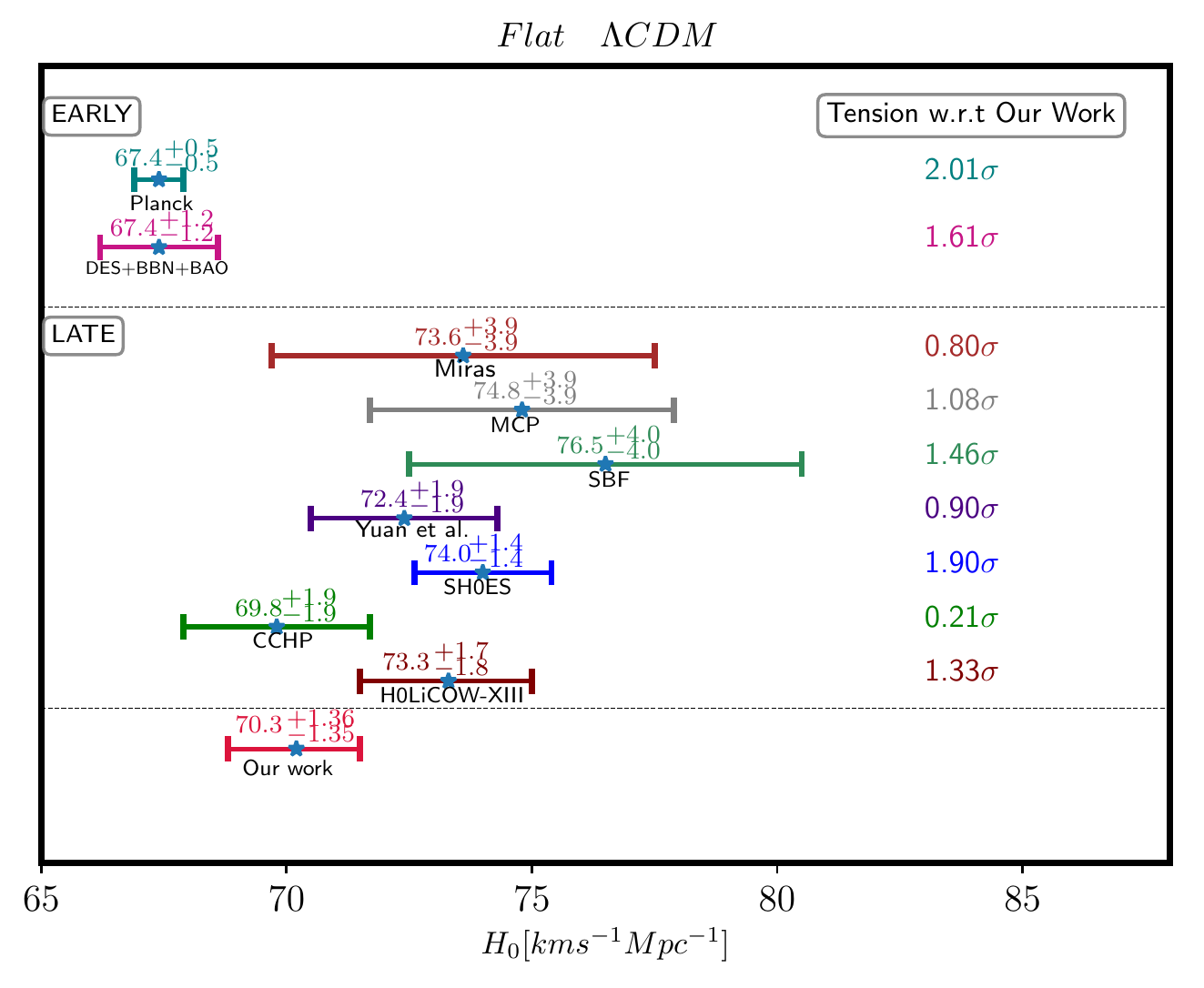}}
\caption{\label{fig:h} $H_{0}$ measurements with $1\sigma$ error bars from different observational data including the one reported in this work for $\Lambda$CDM model. We also show the tensions in our measurement with low-redshift observations compared to other observations. The DES+BBN+BAO, MIRAS, MCP and SBF measurements are taken from \citep{Verde:2019ivm}. For the rest of the measurements, see the text.}
\end{figure}


In Figure \ref{fig:hdr}, we show the constraint in the ($H_{0},r_{d}$) contour plane. As one can see, the constrained contours are very similar for different dark energy models and both $H_{0}$ and $r_{d}$ constraints are consistent with R19 and also with the Planck-2018 measurements.

To summarise, the combination of ``BASE+CC+$f\sigma_{8}$" gives consistent constraints on both $H_{0}$ and $r_{d}$ parameters for different dark energy models showing no preference for models beyond $\Lambda$CDM. Moreover the constraints on $H_{0}= 70.3^{+1.36}_{-1.35}$ for $\Lambda$CDM model is in  $2\sigma$ or less tension with both high-redshift  (Planck-2018, DES+BBN+BAO) and local measurements (R19, F19, H0LiCOW). This is depicted in Figure \ref{fig:h} where different measurements for $H_{0}$ (as quoted in \citep{Verde:2019ivm}) are shown including that from the current analysis. We also show the tension in our current $H_{0}$ measurement with other $H_{0}$ measurements; in all cases the tensions are {within $2\sigma$ or less}.

Finally we calculate the ``Bayesian Evidence" or ``Global Likelihood" for the dark energy models considered in this work. The Bayesian evidence is defined as
\begin{equation}
    {\cal{Z}} = \int P(D|\theta)P(\theta) d^{w}\theta.
\end{equation}

\begin{table}
\centering
\begin{tabular}{ccccc}
\Xhline{\arrayrulewidth}
\textbf{} & \textbf{$\Lambda$CDM} & \textbf{$wCDM$} & \textbf{CPL} & \textbf{PADE}\\ 
\Xhline{\arrayrulewidth}
${Ln {\cal Z}}$ & -60.59 & -69.08 & -63.24 & -63.38 \\
\Xhline{\arrayrulewidth} 
\end{tabular} 
\caption{Bayesian Evidence for different dark energy models.}\label{Table-6}
\end{table}

Here $D$ represents the data considered, $\theta$ represent the model parameters, $w$ is the dimension of the parameter space, $P(D|\theta)$ represents the likelihood function for the parameters $\theta$ given the data $D$ and $P(\theta)$ is the prior distribution for the parameters $\theta$. With this, we use the Jeffrey's scale for Bayesian Evidence to compare models. In particular if $\Delta Ln{\cal Z}$ between two models is more than $2.5$, then the model with higher $Ln {\cal Z}$ is strongly favoured compared to the model with lower $Ln {\cal Z}$ and for  $\Delta Ln{\cal Z} > 5$, it is decisively favoured. In Table \ref{Table-6}, we show the Bayesian evidence for different dark energy models: $\Lambda$CDM has the highest $Ln {\cal Z}$ compared to other dark energy models considered and is therefore still the best model for the low-redshift observational data ``SN+BAO+MASERS+SL+CC+$f\sigma_{8}$". {Interestingly, the $w$CDM is decisively ruled out compared to $\Lambda$CDM by the low-redshift data and both CPL and PADE are also strongly disfavored compared to $\Lambda$CDM.}

\section{Discussions and Conclusions}

\begin{figure}
\resizebox{240pt}{180pt}{\includegraphics{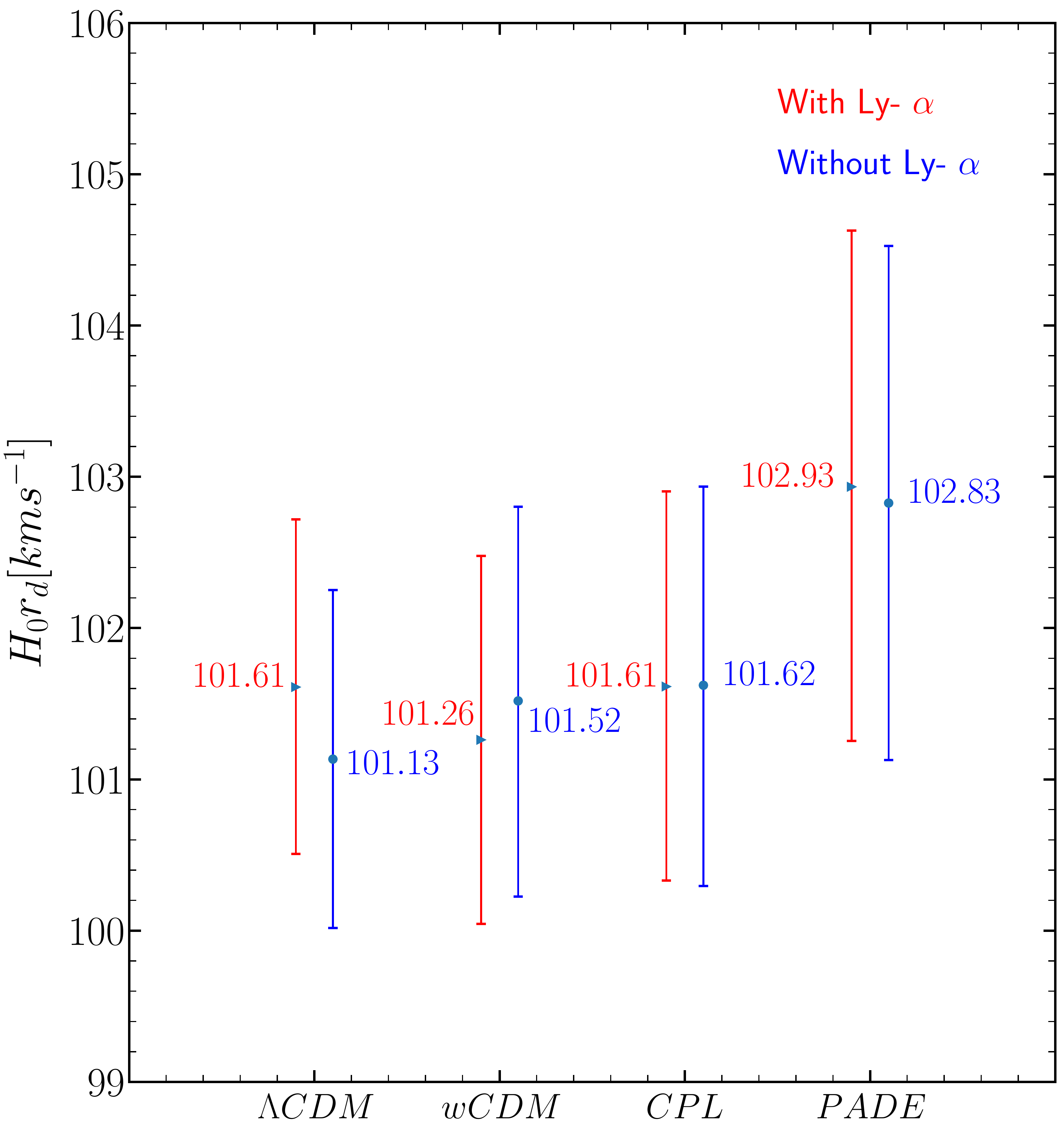}}
\caption{\label{fig3:hdr} Values of $H_{0}r_{d}$ (in Km/s) for different dark energy model. The data set used ``BASE+CC".}
\end{figure}

As discussed, BAO observations measure the combination $H_{0}r_{d}$ rather than $H_{0}$ and $r_{d}$ individually. In our analysis, we include the BAO measurement by BOSS-DR12 with Lyman-$\alpha$ samples at redshift $z=2.4$. To see the effect of this particular observation from BAO-Lyman-$\alpha$ in our analysis, we show in Figure \ref{fig3:hdr}, the constraints on the combination $H_{0}r_{d}$ for different dark energy models with and without this Lyman-$\alpha$ measurement at $z=2.4$. As one can see, the Lyman-$\alpha$ data for BAO at $z=2.4$ does not affect the constraint on the combination $H_{0}r_{d}$. Constraints on $H_{0}r_{d}$ are uniform across different dark energy models considered in this analysis, except for the Pade which gives a slightly higher mean value, and our  $H_{0}r_{d}$ value is consistent with  Planck-2018 within $2\sigma$. There are updated results for BAO-Lyman-$\alpha$ measurement by eBOSS-DR1 \cite{Agathe:2019vsu} for which the tension with Planck-2018 has been reduced. However,  from Figure \ref{fig:hdr} we can safely conclude that incorporating this updated data will have minimal effect on our results.

\begin{figure*}[htp]
  \centering
  {\includegraphics[scale=0.8]{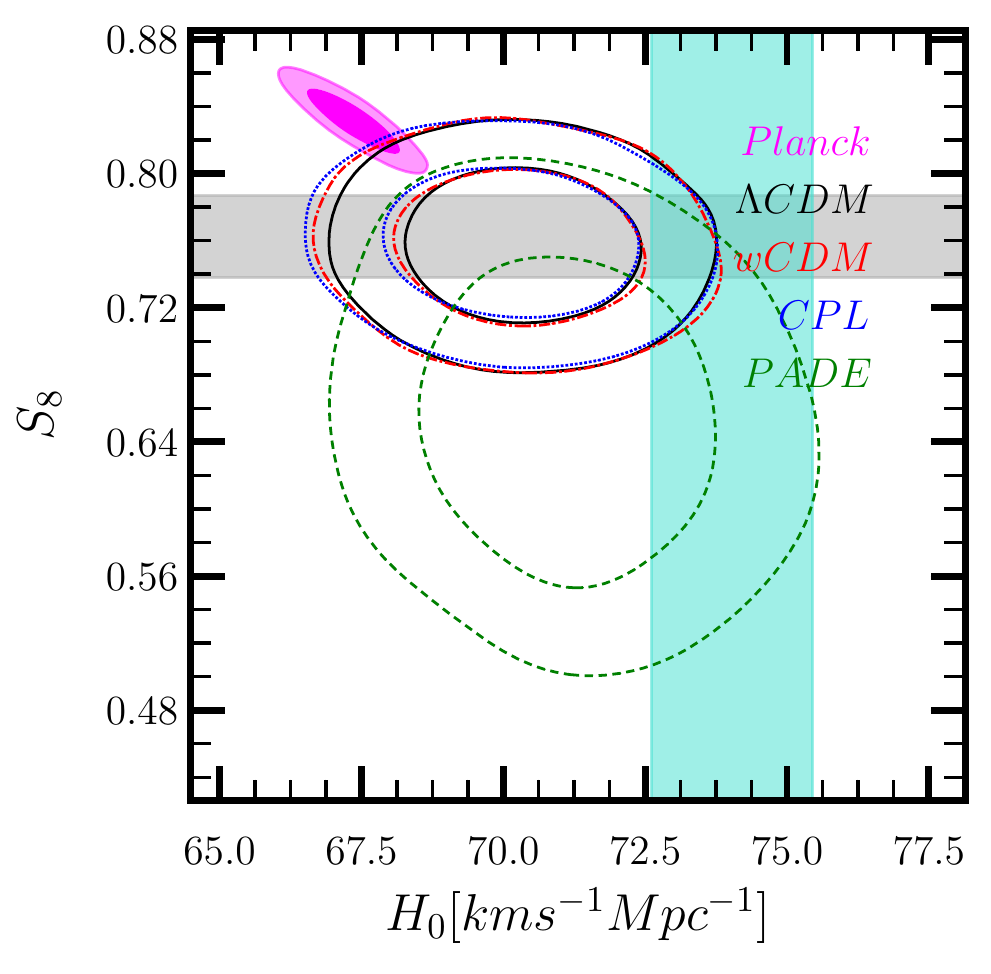}}\qquad\quad
{\includegraphics[scale=0.8]{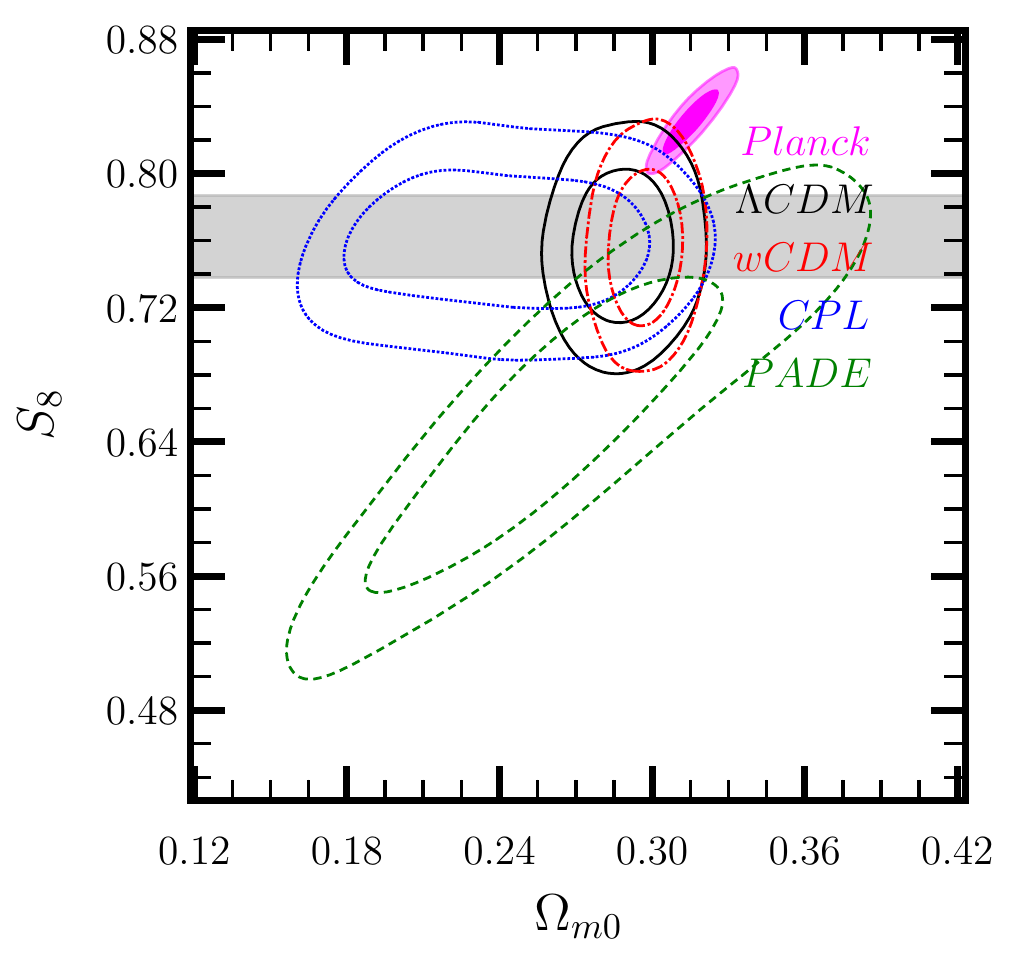}}
\caption{\label{fig:hs8} $1\sigma$ and $2\sigma$ contours in ($H_{0},S_{8}$) plane (Left Figure) and in ($\Omega_{m0},S_{8}$) (Right Figure) for different dark energy models. The grey band is $1\sigma$ bound for $S_{8}$ from KiDS+VIKING-450+DES-Y1 survey results. The green band in the Left Figure is the $1\sigma$ bound on $H_{0}$ from R19. }
\end{figure*}

Next, we explore another tension between the high-redshift CMB  and the corresponding low-redshift measurements. This is related to the growth of large scale structures in the Universe. Weak lensing by large scale structures is one of the most accurate probes to measure the growth of structures in the Universe and the underlying cosmology. The Kilo-Degree Survey (KiDS) spanning $450$ square-degrees in the sky combined with VISTA Kilo Degree Infra-Red Survey (VIKING) \cite{Hildebrandt:2018yau}, has recently performed a robust tomographic cosmic shear analysis  and measured the  $S_{8}$ parameter related to the amplitude of the cosmic shear due to weak lensing: $S_{8} = 0.737^{+0.040}_{-0.036}$. This measured value is in $2.3\sigma$ tension with the corresponding value obtained by Planck-2018 for $\Lambda$CDM. {There is, moreover, a combined tomographic weak lensing analysis of  KiDS+VIKING and Dark Energy Survey (DES-Y1)  \citep{Joudaki:2019pmv}. The constraint on $S_{8}$ from this combined analysis is $S_{8} = 0.762^{+0.025}_{-0.024}$ which is in $2.5\sigma$ tension with Planck-2018 result for $\Lambda$CDM}. These tensions are not as large compared to $H_{0}$ tension discussed before and we still do not know whether this tension between early and late universe measurements is due to unknown systematic or it hints some new physics beyond $\Lambda$CDM model. 

In our current analysis, however, we have obtained an independent constraint on $S_{8}$ solely from the low-redshift observations using measurements of $f\sigma_{8}$ by various surveys. Note that for $\Lambda$CDM model and assuming Einstein GR is the correct theory of gravity, $f\sigma_{8} \sim \Omega_{m}^{0.55}\sigma_{8}$ whereas $S_{8}$ scales as $S_{8} \sim \Omega_{m}^{0.5}\sigma_{8}$. Hence, if $\Lambda$CDM is the true model, using $f\sigma_{8}$ data itself, one also expects to get strong constraint on $S_{8}$. Moreover, if $\Lambda$CDM is a true model, from these two scaling relations, one expects a very small correlation between $S_{8}$ and $\Omega_{m0}$.

For different dark energy models the constraint on $S_{8}$ are shown in Tables \ref{Table-2}, \ref{Table-3}, \ref{Table-4} and \ref{Table-5}. The constraint on $S_{8}$ is uniform over different dark energy models except for Pade where the mean value is substantially less, but with a larger error bar. In particular,  the constraint is  $S_{8} = 0.76^{+0.03}_{-0.03}$ for $\Lambda$CDM which is perfectly consistent with KiDS+VIKING-450+DES-Y1 result, though with slightly larger error bars compared to the latter.   The confidence contours in $(S_{8},\Omega_{m0})$ and in $(S_{8},H_{0})$ parameter space are shown in Figure \ref{fig:hs8} which shows negligible correlation between $S_{8}$ and $\Omega_{m0}$ for $w$CDM and CPL models confirming no deviation from $\Lambda$CDM behaviour.  
This confirms that the independent indirect measurement of $S_{8}$ using low-redshift ``BASE+CC+$f\sigma_{8}$" data set, is fully consistent with the direct measurement of $S_{8}$ by KiDS+VIKING-450+DES-Y1. In this case, too, there is no need to go beyond $\Lambda$CDM.

To conclude, we revisit the low-redshift observational data ``SN+BAO+MASERS+SL+CC+$f\sigma_{8}$" assuming different dark energy models including $\Lambda$CDM. Our goal is to check what the low-redshift data has to say about the current tension in $H_{0}$ measurements from early and late Universe. Our results show that the low-redshift data sets yield consistent constraints for cosmological parameters across different dark energy models and the constraints on $H_{0}$ and $r_{d}$ are consistent with both R19 and Planck-2018 measurements to $2\sigma$ level. In drawing this conclusion, the addition of Cosmic Chronometers data $H(z)$ plays a crucial role by pushing $H_0$ value towards a smaller value (closer to the Planck results), whereas making $r_d$ larger. Also our derived value of $S_{8}$ is fully consistent with the direct measurement of $S_{8}$ by KiDS+VIKING-450+DES-Y1 survey. We show that there is no clear evidence for models beyond $\Lambda$CDM for this set of low-redshift data. Moreover in terms of Bayesian Evidence, $\Lambda$CDM is still the best model compared to other dark energy models analyzed.
Just recently, a improved distance to megamaser NGC-4258 have been measured giving an updated estimate for $H_{0}$ which is $H_{0} = 71.1 \pm 1.9$ KM/s/MPc \citep{Reid:2019tiq}. This is fully consistent with our measured value of $H_{0}$ for $\Lambda$CDM as quoted in Table II with a tension less than $1\sigma$. Moreover, Liao et al have recently obtained a model independent $H_{0}$ constraint using the SnIa and strong lensing measurement by H0LiCOW. Their model independent constraint is $H_{0} = 72.2 \pm 2.1$ KM/s/MPc \citep{Liao:2019qoc} for a spatially flat Universe which is in less than $1\sigma$ tension with our result for $\Lambda$CDM confirming again the consistency of our result with other constraints obtained by different methods.

The $\Lambda$CDM with model parameters as shown in Table \ref{Table-2} is consistent with R19, F19, Planck-2018, H0LiCOW-XIII as well as KiDS+VIKING-450+DES-Y1 and the tensions with each of these observations are always within $2\sigma$ or less. Hence $\Lambda$CDM model with parameters as mentioned in Table \ref{Table-2} best represents the current Universe without any significant tension.

\section*{Acknowledgements}
We are grateful to Ravi Sheth for discussions and Eoin \`O Colg\`ain and Nima Khosravi for comments on the draft. We are also thankful to Adam Riess for his comments on our manuscript.
MMShJ was supported by the  INSF junior chair in black hole physics, grant No 950124 and by Saramadan grant No. ISEF/M/98204. We would like to thank the hospitality of the HECAP section and the Associateship Program of the Abdus-Salam ICTP, Trieste. Ruchika acknowledges the funding from CSIR, Govt. of India under Senior Research Fellowship. Ruchika also acknowledges SISSA for supporting her visit where part of the work has been done.

\vspace{2mm}
\bibliographystyle{apsrev4-1}
\bibliography{new_project}
\end{document}